\input{aipcheck}

\documentclass[
    ,final            
  ]
  {aipproc}

\usepackage{natbib}
\layoutstyle{8x11single}
\usepackage{epstopdf}

\begin{document}

\title{Primordial non-Gaussianity in density fluctuations}


\author{F. Fraschetti}{
  address={\footnotesize Lunar and Planetary Lab \& Dept. Physics, 1629 E University Bvd, University of Arizona, 85721-0092, Tucson, AZ, USA}   
 ,altaddress={LUTh, Observatoire de Paris, CNRS-UMR8102 and Universit\'e Paris VII, 5 Place Jules Janssen, F-92195 Meudon C\'edex, France.} 
}
\author{J.-M. Alimi}{
  address={LUTh, Observatoire de Paris, CNRS-UMR8102 and Universit\'e Paris VII, 5 Place Jules Janssen, F-92195 Meudon C\'edex, France.}
}
\author{J. Courtin}{
  address={LUTh, Observatoire de Paris, CNRS-UMR8102 and Universit\'e Paris VII, 5 Place Jules Janssen, F-92195 Meudon C\'edex, France.}
}
\author{P.-S. Corasaniti}{
  address={LUTh, Observatoire de Paris, CNRS-UMR8102 and Universit\'e Paris VII, 5 Place Jules Janssen, F-92195 Meudon C\'edex, France.}
}


\begin{abstract}
 We present N-body cosmological numerical simulations including a primordial non-Gaussianity
 in the density fluctuation field quantified by the non-linear parameter $f_{NL}$. 
 We have used MPGRAFIC code to produce initial conditions and 
 the Adaptive Mesh Refinement (AMR) code RAMSES to evolve the large scale structure formation.
 We estimated the higher order momenta of the initial distribution of density fluctuations,
 investigated the redshift evolution of the non-linear power spectrum and
 estimated the discrepancy introduced by the
 primordial non-Gaussianity in the non-linear power spectrum.

   \end{abstract}

\maketitle


\section{Introduction}

The most stringent constraints on a primordial non-Gaussianity come
from the temperature anisotropy measurement by WMAP 5-year observations \cite{k09},
which show a small effect of non-Gaussianity on large scales. 
On the other hand, the probe of a positive non-Gaussianity on smaller scales
seems to have been detected recently in WMAP 3-year observations \cite{ym08}.
Extragalactic dust emission can be also associated with primordial deviations from Gaussianity
at sub-degree scales \cite{rsph09}.

From the theoretical point of view, the assumption that primordial density fluctuations
generating large scale structure formation can be modeled by a Gaussian distribution
is supported by the inflationary models. However, distinct variants of the inflationary models
incorporate a primordial non-Gaussianity, as shown by \cite{bkmr04}.

We present a series of N-body numerical simulations of large scale structure formation 
in order to investigate the effect of a primordial non-Gaussianity on the 
probability distribution function and the two-point correlation 
function of the density fluctuation field of the dark matter distribution. 
The non-Gaussianity is quantified by the usual parameter $f_{NL}$,
which modifies the gravitational potential at a primordial epoch.
Higher order modifications of the gravitational potential 
have been recently considered as well \cite{ds09}.

\section{Non-Gaussian model}

For a large class of models of generation of the initial seeds
for structure formation, including single-field \cite{m03} and
multi-field inflation \cite{bu04}, or the curvaton \cite{kk06}, the non-Gaussianity
of the initial density fluctuations can be modeled
through a quadratic term in the Bardeen's gauge-invariant
potential $\Phi$, namely \cite{ks01}
\begin{equation}
\label{FNL}
\Phi (\mathbf{x}) = \Phi_{\rm L} (\mathbf{x}) + f_{\rm{NL}} \left[\Phi_{\rm L}^2 (\mathbf{x}) - 
\langle\Phi_{\rm L}^2 (\mathbf{x})\rangle \right] \;,
\label{NG}
\end{equation}
where $\Phi_{\rm L}$ is a Gaussian random field, 
with $\langle\Phi_{\rm L} (\mathbf{x})\rangle = 0$, and the specific value
of the dimensionless non-linearity parameter $f_{\rm{NL}}$ depends on
the assumed scenario. This form of non-Gaussianity can be obtained
from a truncated expansion of the effective inflaton potential \cite{sb90}.
From the latest WMAP5 temperature data \cite{k09}
\begin{equation}
-151 < f_{\rm{NL}}  < +253 \; \rm with \; 95\% \; CL  
\end{equation}
With this convention positive (negative) $f_{\rm NL}$ corresponds to
positive (negative) skewness of the probability distribution function of density fluctuations.
The Gaussianity tests show that the primordial fluctuations are Gaussian
to the 0.1$\%$ level. 

Since the modification introduced in \eqref{NG} is of second order in the potential,
the linear two-point correlation function is not modified for any value of $f_{NL}$
in the chosen interval $|f_{NL}| \le 5,000$. In fact, by order of magnitude  
$\Phi ({\mathbf x}) \le 10^{-5}-10^{-6}$; so even for $f_{NL} \sim 10^3$,
the non-Gaussian correction gives a contribution of the order of $1\%$.

It should be noted that distinct realizations of primordial non-Gaussianity 
in the density fluctuation field has been proposed, see e.g. \cite{csz07}.

\section{N-body simulations}

The procedure of generation of primordial non-Gaussian initial condition 
has been implemented in the MPI pa\-rallelized numerical code MPGRAFIC \cite{p08}. 
The N-body simulations have been performed using the AMR
parallelized numerical code RAMSES-3.0 \cite{t02}. We have followed only the evolution
of the collisionless dark matter particles.
We performed simulations in computational boxes of physical sizes 
$(162 h^{-1}$ Mpc)$^3$, $(300 h^{-1}$ Mpc)$^3$, 
$(324 h^{-1}$ Mpc)$^3$, $(500 h^{-1}$ Mpc)$^3$ 
with a grid resolution of $256^3$. The series of simulations presented here
aims to detect the non-linear scales effect of the primordial non-Gaussianity.

\section{Cosmological Parameters}

We have adopted the following set of cosmological parameters, within the framework
of the so-called ``concordance'' $\Lambda$-CDM model, compatible with WMAP 5-year observations \cite{k09}:
$\Omega_b = 0.044$, $\Omega_m = 0.26$, $\Omega_\Lambda = 0.74$, $H=72$ km s$^{-1}$Mpc$^{-1}$.
The spectral index of the initial power spectrum is $n=0.951$. 
The rms of density fluctuations in spheres of radius $R_0$, $\sigma^2(R_0)$, is defined as
\begin{equation}
\sigma^2(R_0) = \int d^3 \mathbf{k} P(k;z) W^2(kR_0)
\end{equation}
in absence of non-linear perturbations, i.e. $P(k;z)$ is the linear power spectrum.
Here $W^2(kR_0)$ is an appropriate window function.
At the scale $R_0 = 8 h^{-1}$ Mpc, we imposed $\sigma^2(R_0 )\sim 0.792$ \cite{k09}.

\section{Initial conditions}\label{ini}
 
We first briefly review the procedure of generation of 3D Gaussian initial conditions \cite{b01,p08}. 
In the Gaussian case, the 3D initial density and displacement fields
are completely defined by the two-point correlation function,
or power spectrum, $P(k)$ of the density fluctuation field $\delta({\mathbf x})$, 
defined as $P(k) \delta_D({\mathbf k} - {\mathbf k'}) = \langle \delta({\mathbf k})  ^*\delta({\mathbf k'}) \rangle$,
where $\delta_D({\mathbf x})$ is the Dirac delta function and
\begin{equation}
 \delta({\mathbf k}) = \frac{1}{(2\pi)^3} \int d{\mathbf x} e^{-i {\mathbf k}{\mathbf x}} \delta({\mathbf x})
\label{fourier}
\end{equation}

The 3D density fluctuation field is generated as convolution 
of a Gaussian white noise $n({\mathbf x})$ with the CDM transfer function $T(k) = \sqrt{P(k)}$. 
In the position space, the Gaussian field $n({\mathbf x})$ is generated with the standard technique
of the Gaussian white noise, using a uniform phase 
and an amplitude extracted from the Rayleigh distribution.
The field $n({\mathbf x})$ has zero average value and unitary variance, i.e. $\langle |n({\mathbf x})|^2 \rangle = 1$,
or, equivalently, constant power spectrum: $\langle |n({\mathbf k})|^2 \rangle = 1$.
As a second step, the Gaussian white noise $n({\mathbf x})$ is convolved with a CDM transfer function $T(x)$
to obtain the density fluctuation and displacement fields according to the Zel'dovich approximation \cite{z70}, i.e.
\begin{equation}
\delta({\mathbf x}) = \int d{\mathbf k} e^{i {\mathbf k}{\mathbf x}} n({\mathbf k}) T(k) ,   \quad 
s_j ({\mathbf x}) = \int d{\mathbf k} e^{i {\mathbf k}{\mathbf x}} \frac{n({\mathbf k}) T(k)}{k^2} i k_j.
\label{fourier2}
\end{equation}
with $j=1,2,3$.
Therefore, the primordial fields $\delta({\mathbf x})$ and $s_j ({\mathbf x})$
are made up by two decoupled contributions: the field $n({\mathbf x})$ 
represents the statistics at early epoch, which is commonly assumed to be Gaussian,
providing the phase of the fluctuation. On the other hand, 
the unknown physics of plasma processes leading to the CMB epoch 
is contained in the transfer function $T(k)$, 
completely determined by the two-point correlation function, for which
the parametrization of \cite{eh98,p08} is used. The $T(k)$ provides
the amplitude of the fluctuation and depends on the cosmological model.

In order to disentangle primordial non-Gaussianity 
from non-Gaussianity generated by the non-linear evolution of structures 
under the gravitational potential, the statistics has been modified at early times,
i.e. before the convolution with the transfer function, by using the \eqref{NG}.
Once the Gaussian white noise $n({\mathbf x})$ is generated, 
the primordial density fluctuations field in the Fourier space $\delta^p({\mathbf k}) = n({\mathbf k})$ is related 
through the Poisson equation to the primordial gravitational potential $\Phi^p (\mathbf{x})$:
\begin{equation}
k^2 \Phi^p (\mathbf{k}) = \frac{3 \Omega_{m_0}}{2 a r^2_{H_0}} \delta^p({\mathbf k})
\label{Poisson}
\end{equation}
where $\Omega_{m_0}$, i.e. the matter density in units of critical density, 
and $r_{H_0} = c/H_0$ are evaluated at the present epoch.
The potential $\Phi^p (\mathbf{k})$ is first transformed back to the position space,
and converted in a non-Gaussian potential according to the prescription in \eqref{NG}.
The new gravitational potential $\Phi^p_{NG} (\mathbf{x})$ corresponding to a 
non-Gaussian primordial statistics is converted back to the Fourier space
in which the new density fluctuation and displacement fields are computed.
The initial condition for the N-body simulation is obtained by using the new 
non-Gaussian gravitational potential to perturb the initially uniform density 
distribution according to the Zel'dovich approximation.

The initial redshift, as input for the code RAMSES, is $z_{start} \sim 77$. 

\begin{figure}
 \includegraphics[height=.4\textheight]{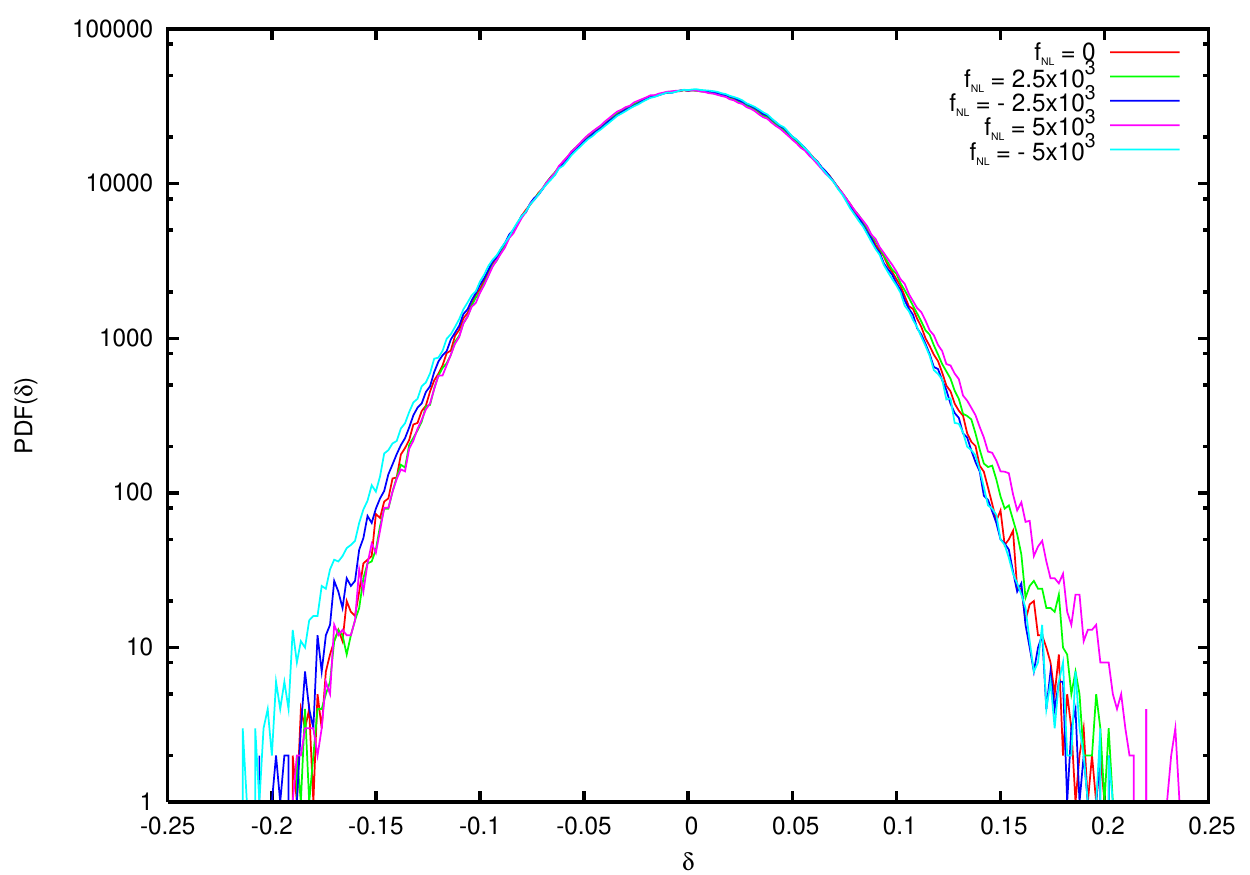}
  \caption{PDF for different values of $f_{NL}$ at redshift $z_{start} \sim 77$.}
  \label{PDF}
\end{figure}

In Fig.~\ref{PDF} the probability distribution function PDF$(\delta)$ is shown for different values of $f_{NL}$. 
The PDF$(\delta)$ of the fluctuation density field $\delta$ at a given scale R is computed 
by sampling the simulation box at  a given redshift with a number of independent spheres of radius R,
fully covering the simulation box.
In Fig.~\ref{PDF} the PDF with comoving sphere radius R $= 8 h^{-1}$ Mpc is shown
in a simulation box of size L = 162 $h^{-1}$ Mpc and initial grid resolution 256$^3$,
with a cell grid comoving size of $\Delta L = 0.63 h^{-1}$ Mpc.

Significant discrepancies with respect to the Gaussian case 
are found in the tail of the PDF only for values of $f_{NL}$
beyond the constraints inferred from WMAP 5-years \cite{k09}, at variance with \cite{g08,p09}.

The second and third order momenta of the $\delta$ distribution 
have been estimated at the initial redshift $z_{start} \sim 77$, corresponding 
to the initial condition existing prior the gravitational evolution under 
the hydrodynamics equations. We find that at $z_{start}$, for 
$f_{NL} = 5000$ and $f_{NL} = -5000$ the third order momentum is 
$\delta^3 = 4.61\times 10^{-8}$ and $\delta^3 = -3.93\times 10^{-8}$, respectively.

As a check the numerical reliability of our computation, 
the third momentum of the non-Gaussian density fluctuation $\delta$ 
can be easily estimated, as pointed out by \cite{ddhs08}. 
From the Laplacian of \eqref{NG},
$\nabla^2  \Phi = \nabla^2 \Phi_{\rm L} + 2f_{\rm{NL}} \left[|\nabla\Phi_{\rm L}|^2 + 
\Phi_{\rm L}\nabla^2\Phi_{\rm L} \right] $,
by using the Poisson equation, the last relation implies
$\langle \delta^3 \rangle = 6 f_{NL} \langle \delta_L^3 \Phi_L \rangle$,
at the lowest order in the product $f_{NL} \Phi_L$. This relation 
has been verified with good accuracy.

\section{Definition of $f_{\rm NL}$}

As already pointed out by several authors, see e.g. \cite{k09,p08},
the definition of $f_{\rm NL}$ depends on the cosmic epoch 
(i.e. on the value of the FRW scale factor $a$) at which \eqref{FNL}
is applied, because both the potentials Gaussian and non-Gaussian, 
$\Phi_{\rm L}$ and $\Phi$ respectively, evolve in time with 
the factor $g(a) = D(a)/a$, where $D(a)$ is the linear growth factor 
of density fluctuations. Therefore the relation between the $f_{{NL}_\infty}$
introduced at $z \rightarrow \infty$ and $f_{{NL}_0}$ introduced at present epoch 
is $f_{{NL}_\infty} = f_{{NL}_0} g(0)/g(\infty)$, where $g(\infty)/g(0) \sim 1.34$.
In the simulations presented here, 
the gravitational potential is modified directly at primordial epoch $z \rightarrow \infty$,
before the convolution with the CDM transfer function $T(k)$. This choice allows
for a direct modification of the primordial statistics (see Sect. above).
Therefore, no linear extrapolation back to $z_{start}$ of the gravitational potential is performed.
The parameter $f_{NL}$ introduced in \eqref{NG} affects only the statistics of the matter distribution and 
does not affect the physics pre-CMB epoch which is included
by the further convolution with the transfer function.

\section{Results}

\begin{figure}
 \includegraphics[height=.27\textheight]{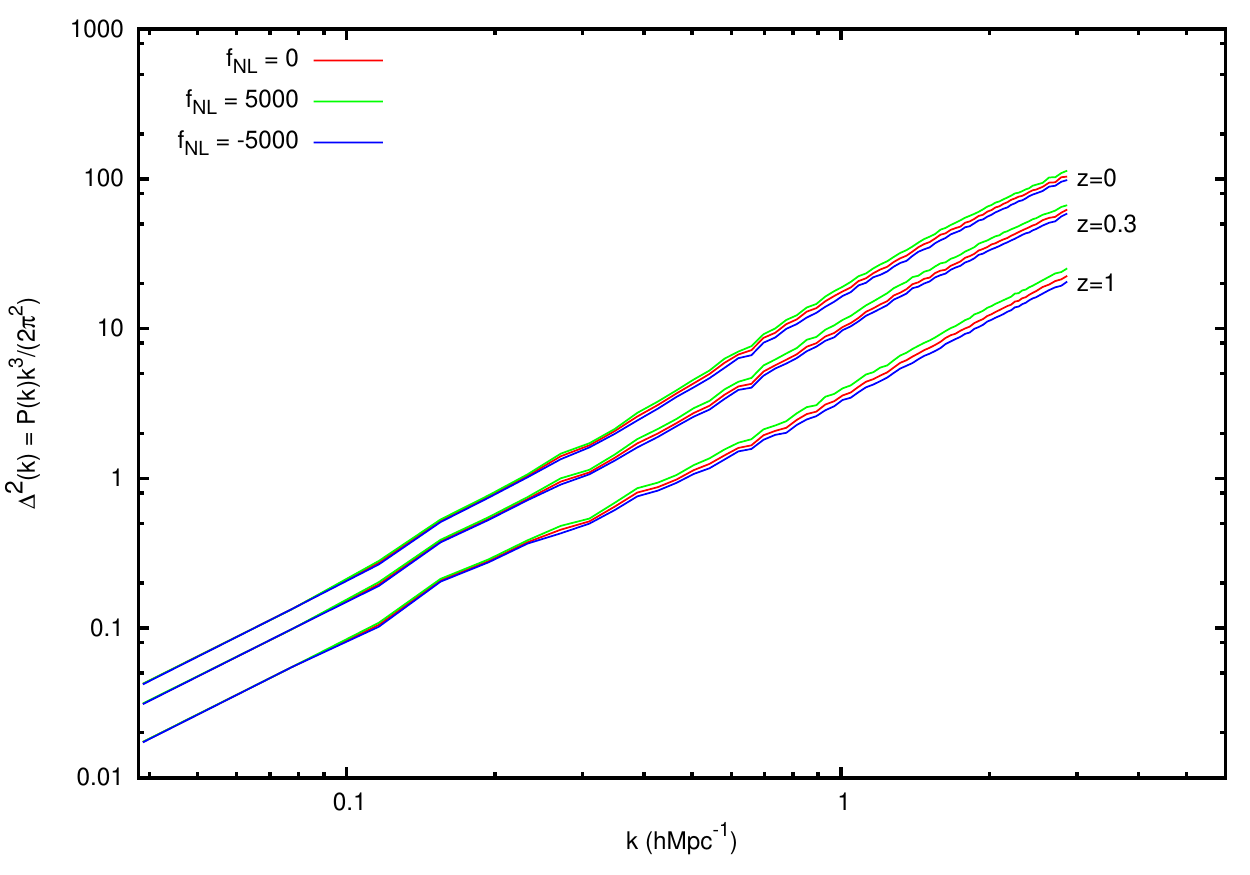}
 \includegraphics[height=.27\textheight]{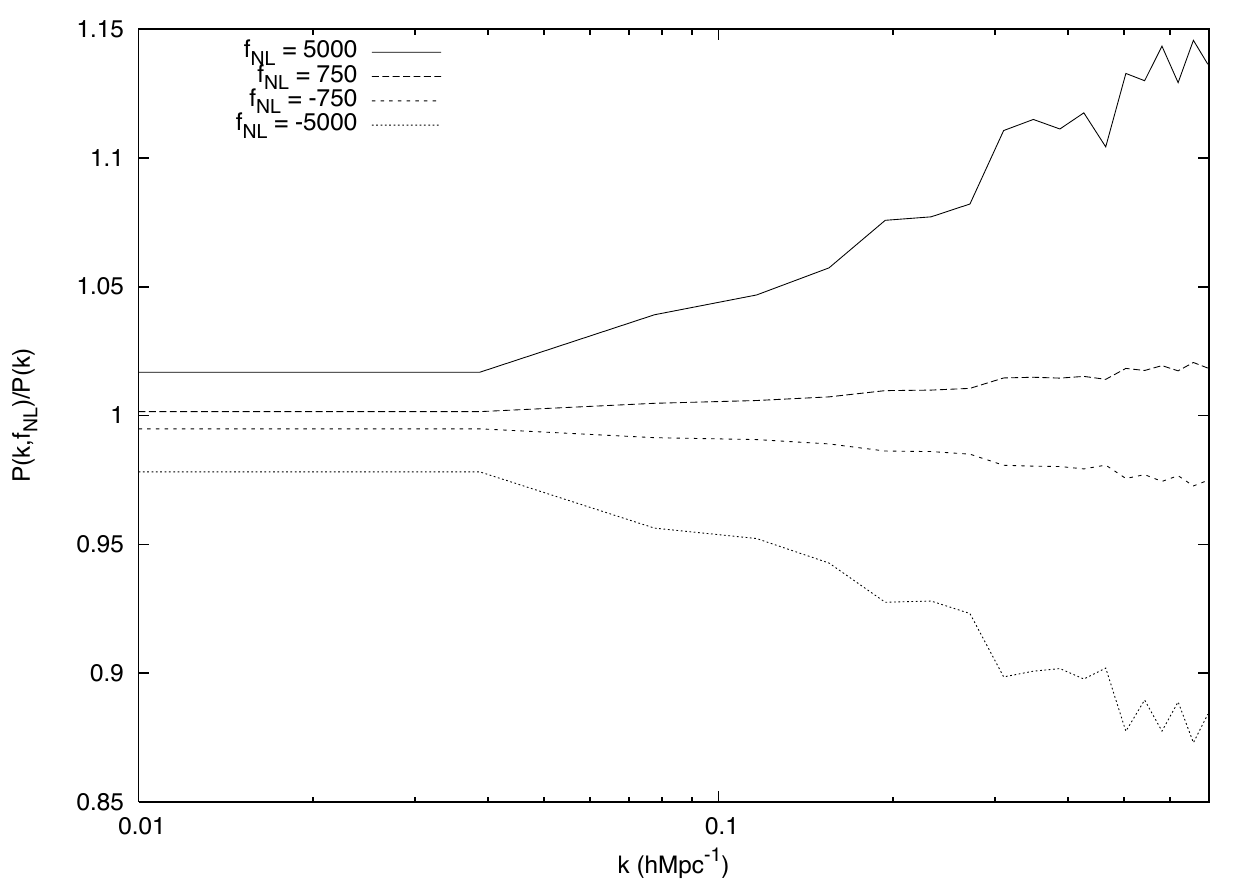}
  \caption{{\it Left panel}: Redshift evolution of the dimensionless power spectrum $\Delta^2(k)$ 
  toward $z=0$, for different values of $f_{NL}$.  {\it Right panel}: 
  Comparison at $z=0$ of the power spectrum of matter density $P(k)$
  in the Gaussian and non-Gaussian case. We find a discrepancy of the
  order of $15\%$ ($2-3\%$) at $k=0.7 (h$ Mpc$^{-1}$), for $f_{NL} = \pm 5000$ ($\pm 750$).}
  \label{Delta}
\end{figure}

 The formalism summarized above allowed us to numerically explore 
       the imprint of the primordial non-Gaussianity of $f_{NL}$-type 
        on the dimensionless power spectrum of the CDM density field $\Delta^2 (k) = P(k)k^3/(2\pi^2)$ at $z=0$,
        see left panel of Fig.~\ref{Delta}.
        The computational box has a size of 162 $h^{-1}$ Mpc, 
        with an initial coarse graining of $256^3$ cells.
        Higher resolution simulations are currently underway.
        The chosen resolution allowed us to resolve the small scale contribution of the non-Gaussian perturbation to the
        structure formation.

  At $z > 5$ the discrepancy between the Gaussian and non-Gaussian 
  case decreases because non-linear effects due to primordial non-Gaussianity 
  on the scale of interest become negligible at large redshifts. 
  
  In the right panel of Fig.~\ref{Delta} the ratio of power spectra $P(k; f_{NL})/P(k; f_{NL} = 0)$ is extracted
from the simulations at redshift $z=0$. The matter power spectrum of non-Gaussian
 models appears to deviate by a few per cent at $k = 0.1 h Mpc^{-1}$, with $f_{NL} = \pm5000$. 
 Moreover, deviations from Gaussian case are enhanced at non-linear scale.


\section{Conclusions and perspectives}
       
The formalism of $f_{NL}$-type allowed us to numerically explore 
the imprint of the primordial non-Gaussianity
on a number of observables quantities, i.e. the initial probability distribution function 
of the density fluctuation field $PDF(\delta)$ and the non-linear power spectrum
at present epoch.

The amplitude of departure of power spectrum from the Gaussian case 
could be hardly detectable, also due to a progressive coalescence 
of power spectra as $z$ decreases.

The detectability of third order momentum of $\delta$, regardeless the 
very small amplitude, can lead to observable differences at the present epoch 
in the case of initial predominance of sub-density regions. 

A series of simulations with higher space resolution will be performed 
to analyse the mass function and the bias of dark matter haloes within the same
model for the primordial non-Gaussianity at the leading perturbative order 
in the gravitational potential, with parameter $f_{NL}$. 
On the other end some simulation tests involving the
successive term in the Taylor expansion of the potential have been carried out,
with the parameter $g_{NL}$ \cite{k09}, leading to not-observable deviations
from Gaussian case in the $PDF$ and $P(k;z=0)$ up to $g_{NL}=10^6$.




\begin{theacknowledgments}
The numerical simulations have been performed with the 
Horizon cluster, for which the technical assistance is gratefully 
acknowledged. The work of FF was supported by CNRS fellowship. 
FF acknowledges the Institut f\"ur Theoretische Physik of Ruhr-Universit\"at Bochum,
where part of this work has been done.  
\end{theacknowledgments}



\bibliographystyle{aipproc}   



\end{document}